\documentclass[lettersize,journal]{IEEEtran}

\usepackage[T1]{fontenc}
\usepackage{microtype}
\usepackage{amsmath}
\usepackage{graphicx}
\usepackage{eso-pic}
\usepackage[caption=false,font=normalsize,labelfont=sf,textfont=sf]{subfig}
\usepackage{placeins}
\usepackage{capt-of}
\usepackage{booktabs}
\usepackage[normalem]{ulem}
\usepackage{tabularx}
\usepackage[skins]{tcolorbox}
\usepackage{xurl}
\usepackage{cite}
\usepackage[hidelinks]{hyperref}

\newcommand{\tool}{Claude Code}
\newcommand{\summarybox}[2]{%
  \medskip
  \begin{tcolorbox}[enhanced,title=#1,
    attach boxed title to top left={xshift=0mm,yshift*=-\tcboxedtitleheight/2}]
    #2
  \end{tcolorbox}
}

\usepackage{xcolor}
\usepackage{xspace}
\usepackage{pifont}

\newcommand{\circlednum}[1]{\ding{\numexpr171+#1\relax}}

\newcommand{\rqone}{How accurately can \tool{} identify BICs without the code-tracing assumption?}
\newcommand{\rqtwo}{What search strategies does \tool{} use to identify BICs?}
\newcommand{\rqthree}{Why does \tool{} fail to identify BICs correctly?}

\usepackage{framed}
{\begin{minipage}{0.95\columnwidth}%
		}
		{\end{minipage}}%
\usepackage{xparse}

\newsavebox{\fminipagebox}
\NewDocumentEnvironment{hassanbox}{m O{\fboxsep}}
{\par\kern#2\noindent\begin{lrbox}{\fminipagebox}
		\begin{minipage}{#1}\ignorespaces}
			{\end{minipage}\end{lrbox}%
	\makebox[#1]{%
		\kern\dimexpr-\fboxsep-\fboxrule\relax
		\fbox{\usebox{\fminipagebox}}%
		\kern\dimexpr-\fboxsep-\fboxrule\relax
	}\par\kern#2
}

{\begin{center}\vspace{-0mm}\noindent\begin{Sbox}\begin{minipage}{0.95\columnwidth}}%
				{\end{minipage}\end{Sbox}\fbox{\TheSbox}\end{center}\vspace{0mm}}

\title{Rethinking SZZ Search Strategies: \\Lessons from Claude Code}

\author{Inase~Kondo, Masanari~Kondo, Daniel~M.~German,
        Olivier~Nourry, Yasutaka~Kamei, and~Yoshiki~Higo%
\thanks{I. Kondo, O. Nourry, and Y. Higo are with the
Department of Computer Science, Graduate School of Information Science
and Technology, The University of Osaka, Japan.}%
\thanks{M. Kondo and Y. Kamei are with the Faculty of
Information Science and Electrical Engineering, Kyushu University, Japan.}%
\thanks{D. M. German is with the Department of Computer
Science, University of Victoria, Canada.}}

\begin{document}

\AddToShipoutPictureFG*{%
  \AtPageLowerLeft{%
    \put(\LenToUnit{0.5\paperwidth},18){%
      \makebox[0pt][c]{%
        \parbox[b]{\textwidth}{\centering\fontsize{7}{8.5}\selectfont
          This work has been submitted to the IEEE for possible publication.
          Copyright may be transferred without notice, after which this version
          may no longer be accessible.}%
      }%
    }%
  }%
}

\bstctlcite{IEEEBSTcontrol}

\maketitle
\begin{abstract}
	Identifying bug-inducing commits (BICs) is important for studying software bugs and maintaining software, yet the accuracy of existing SZZ methods remains limited.
	Although various improvements have been proposed, many SZZ methods still rely on code tracing, which searches for BICs by tracing code changed in a bug-fixing commit (BFC) back to earlier commits.
	However, this approach may miss the actual BIC when the code responsible for introducing a bug differs from the code modified to fix it.
	To overcome this limitation, we investigate whether BIC identification can be improved by exploring search strategies that are not constrained to start with code tracing.
	We use Claude Code, a coding agent that has demonstrated high BIC identification accuracy in prior research, to explore such strategies.
	Rather than prescribing a search process or providing predefined candidate commits, we allow Claude Code to independently investigate a Git repository.
	We evaluate its identification accuracy and search behavior on 79 Java cases from 70 repositories.
	In 79.7\% of runs, Claude Code reviewed commit histories before inspecting individual code diffs.
	It used commit messages to select changes for closer inspection.
	With this approach, Claude Code achieves an F1 score of 0.771, significantly outperforming all ten evaluated SZZ baselines.
	Our failure analysis also reveals incorrect oracle labels and ambiguities in BIC definitions in the studied evaluation dataset.
	After reconstructing the evaluation dataset to address these issues, Claude Code achieves an F1 score of 0.938.
	A key lesson from Claude Code for improving SZZ search strategies is to use semantic information to find relevant commits and then examine their code changes, rather than starting with code tracing.
\end{abstract}


\section{Introduction}
\label{sec:introduction}
\IEEEPARstart{B}{ug-inducing} commits (BICs) are important for understanding bugs and predicting defects in both software development and research~\cite{kim2008classifying,kamei2013jit,mcintosh2018moving,pornprasit2021jitline,vandehei2021leveraging,bao2022vszz,bernardi2018communication,linuxSubmittingPatches,li2024patchporting}.
The SZZ method is widely used to identify BICs automatically~\cite{sliwerski2005changes}.
However, its accuracy remains limited~\cite{rosa2023comprehensive,lyu2024evaluating}.
To improve its accuracy, researchers have proposed rule-based and, more recently, LLM- and agent-based SZZ methods~\cite{kim2006automatic,davies2014comparing,dacosta2017framework,neto2018impact,sahal2018additions,perezrosero2024wia,kondo2026mvszz,tang2025llm4szz,shi2026agenticszz,lyu2026agentszz,risse2026agents}.
Despite these advances, there is still room to improve BIC identification accuracy.

One limitation of many existing SZZ methods is their reliance on \emph{code tracing}.
In this search process, the method first uses information from a bug-fixing commit (BFC) to select code that may have caused the bug.
It then uses \texttt{git blame} to find earlier commits that changed this code.
In other words, this method limits the search space to BICs that include code that can be traced directly from the BFC, potentially missing relevant changes elsewhere in the codebase.
The BIC and BFC sometimes change different functions or files, making this inference difficult~\cite{wen2019exploring,rosa2023comprehensive,lyu2024evaluating}.

LLMs use semantic information about changes to guide their search.
Developers use commit messages to describe what was changed and why~\cite{tian2022good,li2023commit,dhaouadi2024rationale}.
These messages may help LLMs find commits that are relevant to the bug.
However, the previous LLM- and agent-based SZZ methods use semantic information only to narrow the search space for code tracing, rather than as the primary means of identifying BICs.

In this study, we question the common assumption in SZZ methods that the search should start with code tracing.
Our main hypothesis is that LLMs can identify BICs more accurately if they are not constrained to start with code tracing.
Hence, we explore alternative search processes that do not start with code tracing to improve BIC identification accuracy.
We use Claude Code, which achieved high BIC identification accuracy in prior work on Simple-SZZ-Agent~\cite{risse2026agents}.
We give Claude Code a Git repository, BFC information, and a simple prompt asking it to identify BICs, without prescribing a search process.
We investigate how accurately Claude Code can identify BICs without this assumption and which search processes may improve identification accuracy.
We also analyze why Claude Code fails to identify BICs under these conditions.

We evaluate Claude Code on 79 bug-fixing commits involving Java code from 70 repositories in Rosa et al.'s Developer-Informed Oracle Dataset~\cite{rosa2023comprehensive}.
For each BFC, the dataset provides ground-truth BICs (the oracle).
To reduce the risk of answer leakage, we removed explicit BIC information from BFC commit messages and blocked Internet access from Claude Code's investigation tools.
We then gave Claude Code the repository and BFC information and ran it three times for each BFC.
We address the following research questions:

\begin{enumerate}
	\item \textbf{RQ1: Identification Accuracy.} \rqone{}
	\item \textbf{RQ2: Search Strategies.} \rqtwo{}
	\item \textbf{RQ3: Failure Causes.} \rqthree{}
\end{enumerate}

For RQ1, \tool{} achieved an F1 score of 0.771, outperforming all evaluated baselines, including Simple-SZZ-Agent, which achieved the second-highest F1 score of 0.696.
For RQ2, \tool{} first reviewed a list of commits and then examined the diffs of individual commits in 79.7\% of runs.
In 54.4\% of runs, it began with \texttt{git log -{}-oneline}, which lists commit hashes and message subjects without code diffs.
For RQ3, our analysis revealed not only mistakes by \tool{} but also incorrect oracle labels and ambiguity in the definition of a BIC.
These results suggest a promising new direction for SZZ methods: using information such as commit messages to search history for relevant changes and then examining commits in detail.

This study makes the following three contributions:

\begin{itemize}
	\item We show that \tool{} achieves higher BIC identification accuracy than all evaluated baselines. It does so without SZZ-specific investigation instructions, specialized tools, or prepared candidate sets.
	\item We characterize a BIC investigation process that is not limited to code tracing. \tool{} reviews commit histories, including commit messages, narrows down candidate commits, and examines their changes.
	\item We identify incorrect oracle labels, ambiguities in BIC definitions, and BIC information leakage from BFC messages.
	      We address these issues and provide a revised dataset\footnote{\url{https://doi.org/10.5281/zenodo.23271971}} better suited to evaluating LLM- and agent-based SZZ methods.
\end{itemize}

\section{Background}
\label{sec:background}

\subsection{Code Tracing in SZZ}
\label{sec:bg-assumptions}

The basic SZZ method uses \texttt{git blame} to trace lines deleted or modified by a BFC and find the commits that last changed them~\cite{sliwerski2005changes}.
However, the resulting commits may contain refactorings or other changes that did not introduce the bug~\cite{kim2006automatic,dacosta2017framework,neto2018impact}.
A BFC may also contain changes unrelated to the bug, and tracing these lines can produce unrelated candidates~\cite{herzig2013tangled,rani2024discussion}.
Existing studies have addressed these problems to improve BIC identification accuracy.

Rule-based SZZ methods mainly focus on selecting code to trace and selecting BICs from the resulting candidates~\cite{kim2006automatic,dacosta2017framework,neto2018impact,davies2014comparing,sahal2018additions,perezrosero2024wia,kondo2026mvszz,watanabe2024token}.
AG-SZZ uses an annotation graph to trace lines across revisions while excluding cosmetic changes, such as changes to whitespace~\cite{kim2006automatic}.
MA-SZZ accounts for merges and other history operations during tracing~\cite{dacosta2017framework}.
RA-SZZ excludes refactoring-related changes from tracing~\cite{neto2018impact}.
Among the resulting candidates, R-SZZ selects the most recent commit, whereas L-SZZ selects the commit with the largest number of changed lines~\cite{davies2014comparing}.

LLM- and agent-based SZZ methods use code semantics to select relevant code for tracing and assess candidate commits~\cite{tang2025llm4szz,shi2026agenticszz,lyu2026agentszz,risse2026agents}.
LLM4SZZ selects bug-related statements from the code around a fix and traces their origins~\cite{tang2025llm4szz}.
It then compares the code in each candidate commit with the code in its parent to assess whether the candidate introduced the bug.
AgentSZZ also recommends starting with \texttt{git blame} on lines modified by the fix~\cite{lyu2026agentszz}.
If this does not identify the BIC, it allows other searches, such as examining changes to relevant strings or functions.

These methods have improved BIC identification accuracy.
However, many still share the same basic approach: select suspicious code and trace its history to find the commit that introduced it.

\subsection{Limitations of Code Tracing}
\label{sec:bg-tracing-limitations}

Many SZZ methods rely on code tracing, but this approach has two main limitations.

First, the code changed to fix a bug may differ from the code that introduced it~\cite{wen2019exploring,rosa2023comprehensive,lyu2024evaluating,rodriguezperez2020bugs}.
Lyu et al. studied BICs in the Linux kernel.
For more than 13\% of these BICs, the BIC and its BFC changed completely different files~\cite{lyu2024evaluating}.
For example, a developer may rename a function but forget to update the code that calls it~\cite{wen2022quick}.
In this case, the BIC renames the function, and the BFC fixes the call.
Tracing the history of the calling code alone does not find the commit that renamed the function.

Second, searching only the commits found by code tracing may miss other related commits~\cite{rezk_ghost_2022}.
Commit messages can describe what changed and why~\cite{tian2022good,li2023commit,dhaouadi2024rationale}.
These messages may help find commits related to the same feature, even if they change different code.
Using messages to find new candidates, not just to check existing ones, may help find commits missed by code tracing~\cite{rosa2023comprehensive}.

\subsection{Investigating BICs with Claude Code}
\label{sec:bg-search-process}

Risse and Böhme showed that SZZ-Agent and Simple-SZZ-Agent, both using Claude Code, can achieve high BIC identification accuracy~\cite{risse2026agents}.
Claude Code has also been evaluated on terminal tasks, software development, and vulnerability repair~\cite{merrill2026terminalbench,xu2025swecompass,zhang2025bountybench}.
A GitHub study found that most Claude Code-assisted pull requests were merged, sometimes after human revisions~\cite{watanabe2026agentic}.
These results suggest that Claude Code may identify BICs by deciding what to investigate and which operations to perform.

SZZ-Agent first asks the agent to assess candidates generated by \texttt{git blame}.
If no candidate is selected, it searches the histories of files modified by the BFC using agent-guided binary search.
Simple-SZZ-Agent removes this initial stage and binary search, letting the agent investigate and select from the collected candidates.
The authors reported that Simple-SZZ-Agent matched or exceeded SZZ-Agent’s F1 score on all four datasets.
These results show that high identification accuracy can be achieved without prescribing these search steps.

However, Simple-SZZ-Agent still provides candidate commits with code tracing information and investigation instructions in advance.
It gives Claude Code a folder of candidate commit diffs rather than a Git repository.
These candidates are collected from the histories of files modified by the BFC.
Its prompt asks Claude Code to understand the bug, identify the problematic code, and select the commit that first introduced that code.
This setup retains the code-tracing perspective and limits the search space through SZZ-specific preparation.

We analyze \tool{}'s search process to identify alternatives to code tracing for finding BICs.
Unlike Simple-SZZ-Agent, we do not provide predefined candidate sets or SZZ-specific search instructions.
Instead, we give \tool{} only a Git repository, BFC information, and a simple prompt asking it to identify BICs.
\tool{} chooses what to investigate and which commands to execute through its built-in Bash tool.
We evaluate its BIC identification accuracy and examine how it uses the available information to find candidates.

\section{BIC Identification with Claude Code}
\label{sec:approach}

We ask \tool{} to identify BICs from a repository and BFC information.
This section describes the settings, repository preparation, investigation prompt, and communication logging used in our study.

\subsection{Claude Code Settings}
\label{subsec:execution-settings}

We use \tool{} 2.1.232~\cite{anthropicClaudeCode} with Claude Sonnet 5~\cite{anthropic2026sonnet5} and \texttt{high} reasoning effort.
Table~\ref{tab:claude-settings} summarizes the settings.

\par\addvspace{\intextsep}
\noindent\begin{minipage}{\columnwidth}
	\centering
	\captionof{table}{Claude Code settings.}
	\label{tab:claude-settings}
	\small
	\setlength{\tabcolsep}{2pt}
	\begin{tabular*}{\linewidth}{@{\extracolsep{\fill}}ll@{}}
		\toprule
		Setting & Value \\
		\midrule
		Version & 2.1.232 \\
		Model / effort & Claude Sonnet 5 / high \\
		Tools & Bash only \\
		Permissions & \texttt{dontAsk}; read-only repository \\
		Time limit & 600 s per invocation \\
		Cost threshold & USD 5 per attempt; no cumulative cap \\
		User/project settings & Disabled \\
		Memory / MCP & Disabled \\
		Initial Git context & Status and 5 recent commits \\
		External access & Web tools and network disabled \\
		API capture & Request and response bodies \\
		\bottomrule
	\end{tabular*}
\end{minipage}
\par\addvspace{\intextsep}

We enable only the built-in Bash tool for Git searches and file access.
Without SZZ-specific tools, \tool{} chooses its commands, arguments, and search targets.

To reduce the risk of answer leakage, we remove BIC-identifying information from the BFC commit message before providing the repository to \tool{}.
We make the repository read-only to prevent changes during the investigation.
We also prohibit external searches, delegation to additional agents, and access outside the working area.

Before the first tool call, \tool{} automatically collects Git status and summaries of the five latest commits.
We retain this initial Git context without modification.
Each investigation starts in a new session without previous conversations.
We disable user and project settings and automatic memory to avoid personalized context.

For each attempt, we set a time limit of 600 seconds and a stopping threshold of USD~5 in reported cost.
These limits keep the resources spent on a single investigation within a practical budget.
The Claude Code interface used in this study does not allow us to set a random seed or temperature.
We therefore run each case three times and report the results.

\subsection{Repository Preparation}
\label{subsec:repository-preparation}

\begin{figure}[!t]
\centering
  \includegraphics[width=\linewidth]{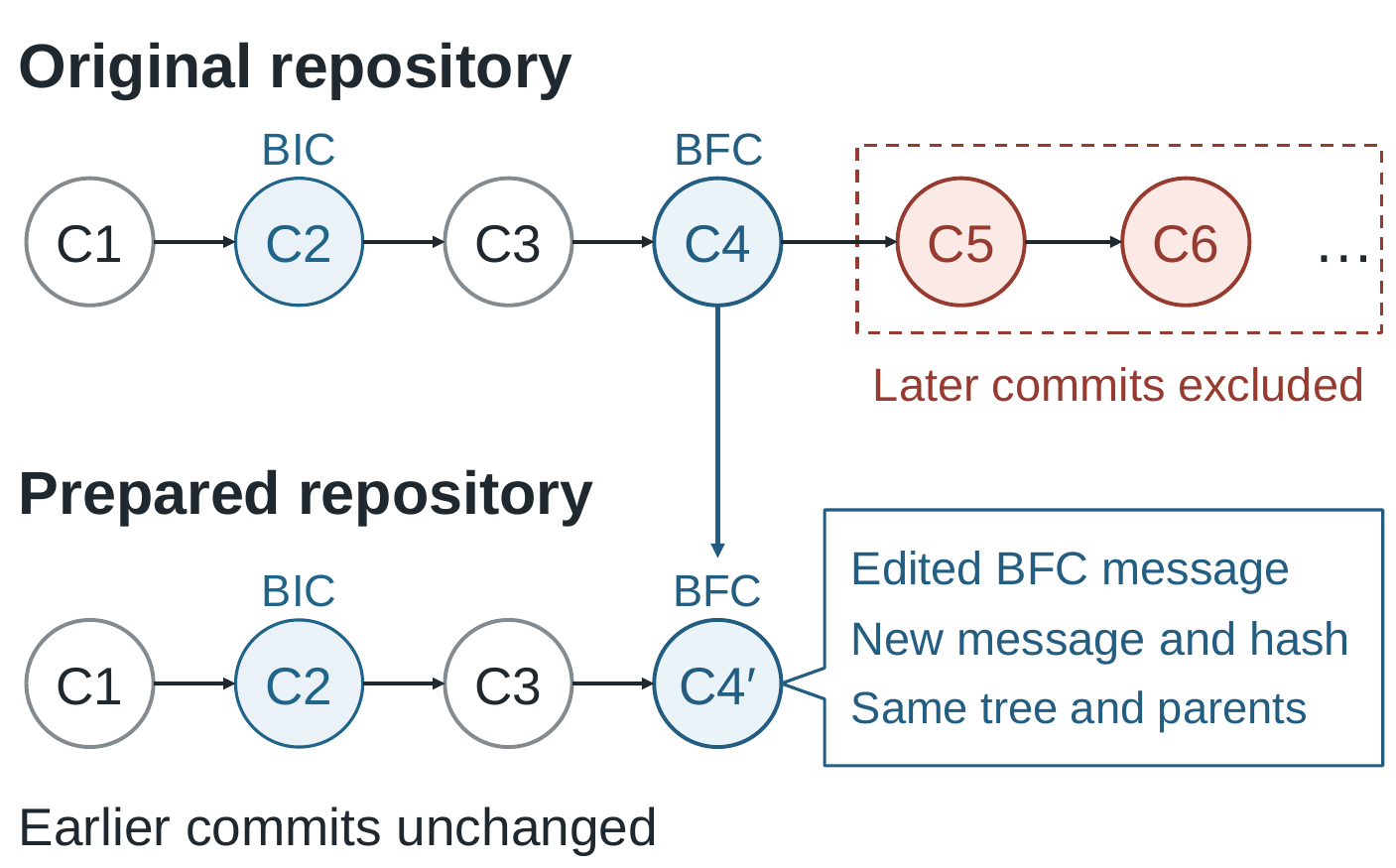}
\caption{Repository preparation: replace BFC C4 with its message-edited version C4$'$ and exclude later commits.}
\label{fig:repository-preparation}
\end{figure}

Figure~\ref{fig:repository-preparation} shows how we prepare a repository for each BFC.
We create a separate repository containing the BFC and all its ancestor commits.
This prevents access to information added after the fix.
We then remove BIC-identifying information from the BFC message and recreate the BFC with the edited message.
Section~\ref{subsec:message-preparation} describes this editing process.
In Figure~\ref{fig:repository-preparation}, C4 denotes the original BFC and C4$'$ denotes its replacement.
C4$'$ preserves the file snapshot and parent commits of C4.
We provide the hash of C4$'$ to \tool{} as the BFC hash.

\subsection{BFC Message Preparation}
\label{subsec:message-preparation}

We evaluate \tool{} on Rosa et al.'s Developer-Informed Oracle Dataset~\cite{rosa2023comprehensive}.
This dataset has also been used to evaluate LLM4SZZ, AgenticSZZ, AgentSZZ, SZZ-Agent, and Simple-SZZ-Agent~\cite{tang2025llm4szz,shi2026agenticszz,lyu2026agentszz,risse2026agents}.
Its BFC messages contain developers' explicit references to BICs.

Prior studies mask BIC hashes to prevent answer leakage~\cite{shi2026agenticszz,lyu2026agentszz,risse2026agents}.
However, we found that the original messages also contain other information that identifies BICs.
For example, a description of a feature added by the BIC can narrow the search to the commit that added it.
SZZ is needed when the introducing commit is unknown.
We aim to evaluate whether \tool{} can identify the BIC from repository history, rather than read an answer stated in the BFC message.
We therefore remove BIC-identifying information while retaining the bug symptoms and fix information.

We first analyzed the original messages of all 79 cases through open coding.
Two authors independently examined the messages and discussed their labels.
All 79 messages contained a BIC hash, and 32 also contained other BIC-identifying information.
Table~\ref{tab:bic-message-information} groups this information into four overlapping categories: change descriptions, linked issues or PRs, authors, and dates.
BIC hashes directly identify the answer, whereas the other information can narrow the search.

\begin{table}[!t]
	\centering
	\caption{BIC information in original BFC messages and its leakage risks.}
	\label{tab:bic-message-information}
	\footnotesize
	\setlength{\tabcolsep}{3pt}
	\renewcommand{\arraystretch}{1.12}
	\begin{tabularx}{\columnwidth}{@{}l>{\raggedright\arraybackslash}Xr@{}}
		\toprule
		Category                                          & Leakage risk                         & Cases \\
		\midrule
		Hash                                              & Directly identifies the BIC          & 79    \\
		\midrule
		\multicolumn{2}{@{}l}{\textbf{Other Information}} & \textbf{32}                                  \\
		\quad Change Description                          & Points to the change to search for   & 17    \\
		\quad Linked Issue or PR                          & Points to linked commits             & 11    \\
		\quad Author                                      & Narrows the search by author         & 6     \\
		\quad Date                                        & Narrows the search by date or period & 5     \\
		\bottomrule
	\end{tabularx}
	\par\smallskip
	\begin{minipage}{\linewidth}
		\footnotesize
		The four categories under Other Information overlap.
	\end{minipage}
\end{table}

We deleted BIC hashes and other identifying information wherever appropriate.
We reworded text minimally only where deletion would break the grammar or alter its meaning.
One author edited the messages, and another reviewed them.
They disagreed on one message and resolved the disagreement through discussion.

\begin{figure}[!t]
  \centering
  \includegraphics[width=\columnwidth]{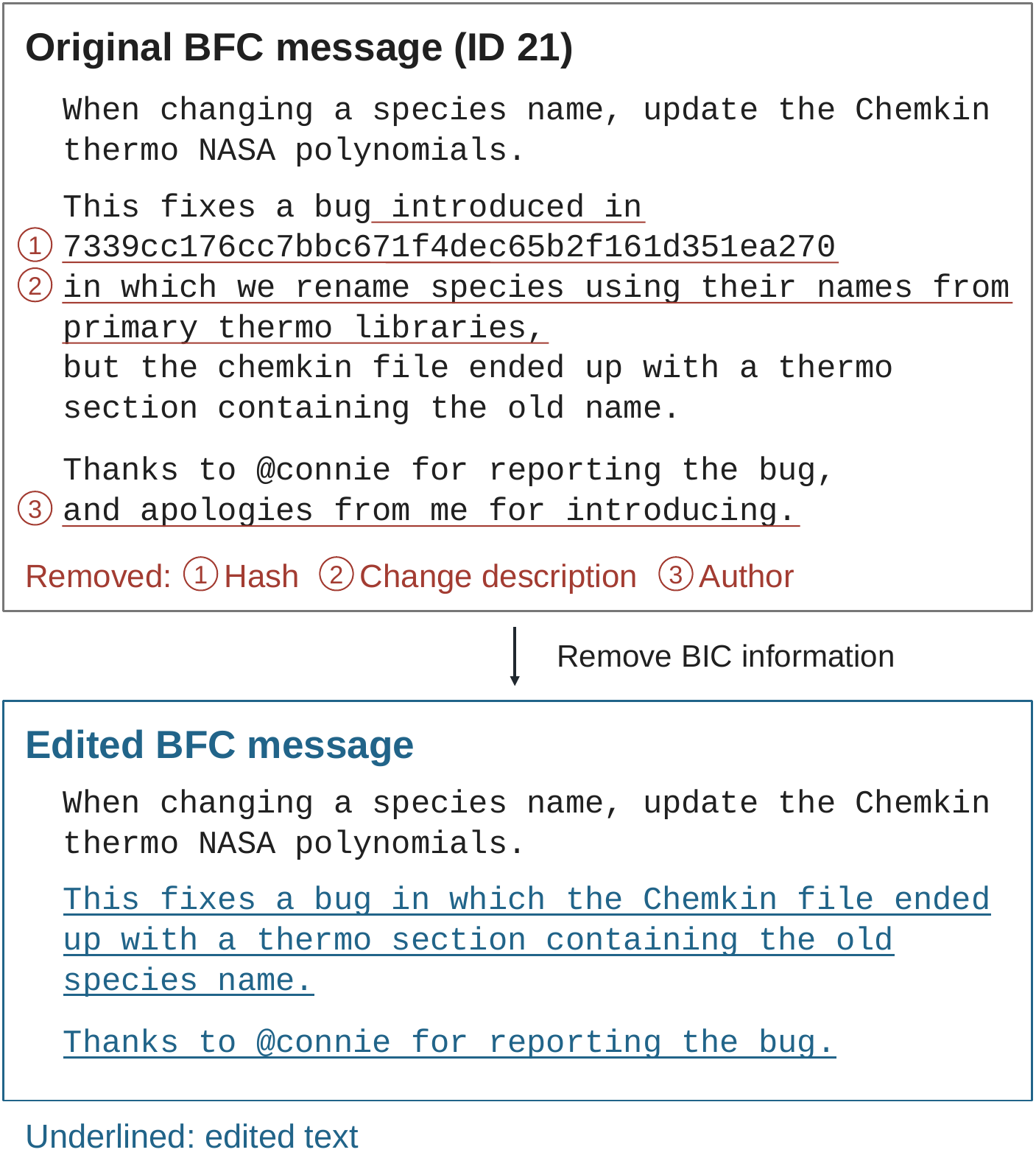}
  \caption{BFC message editing for case 21. Numbered underlines identify removed BIC information; symptoms and fix details remain.}
  \label{fig:message-neutralization}
\end{figure}

Figure~\ref{fig:message-neutralization} shows the edit for case~21.
The hash marked \circlednum{1} directly identifies the BIC.
The species-renaming description marked \circlednum{2} can narrow the search to earlier commits that renamed the species.
The author reference \texttt{me} marked \circlednum{3} identifies the message writer as the BIC author, narrowing the search to that person's changes.
We removed these clues while retaining the symptom---the old species name remaining in the file---and the description of the fix.
The replication package includes the edited BFC commit messages.

\subsection{Investigation Prompt}
\label{subsec:investigation-prompt}

\begin{figure}[!t]
  \centering
  \includegraphics[width=\columnwidth]{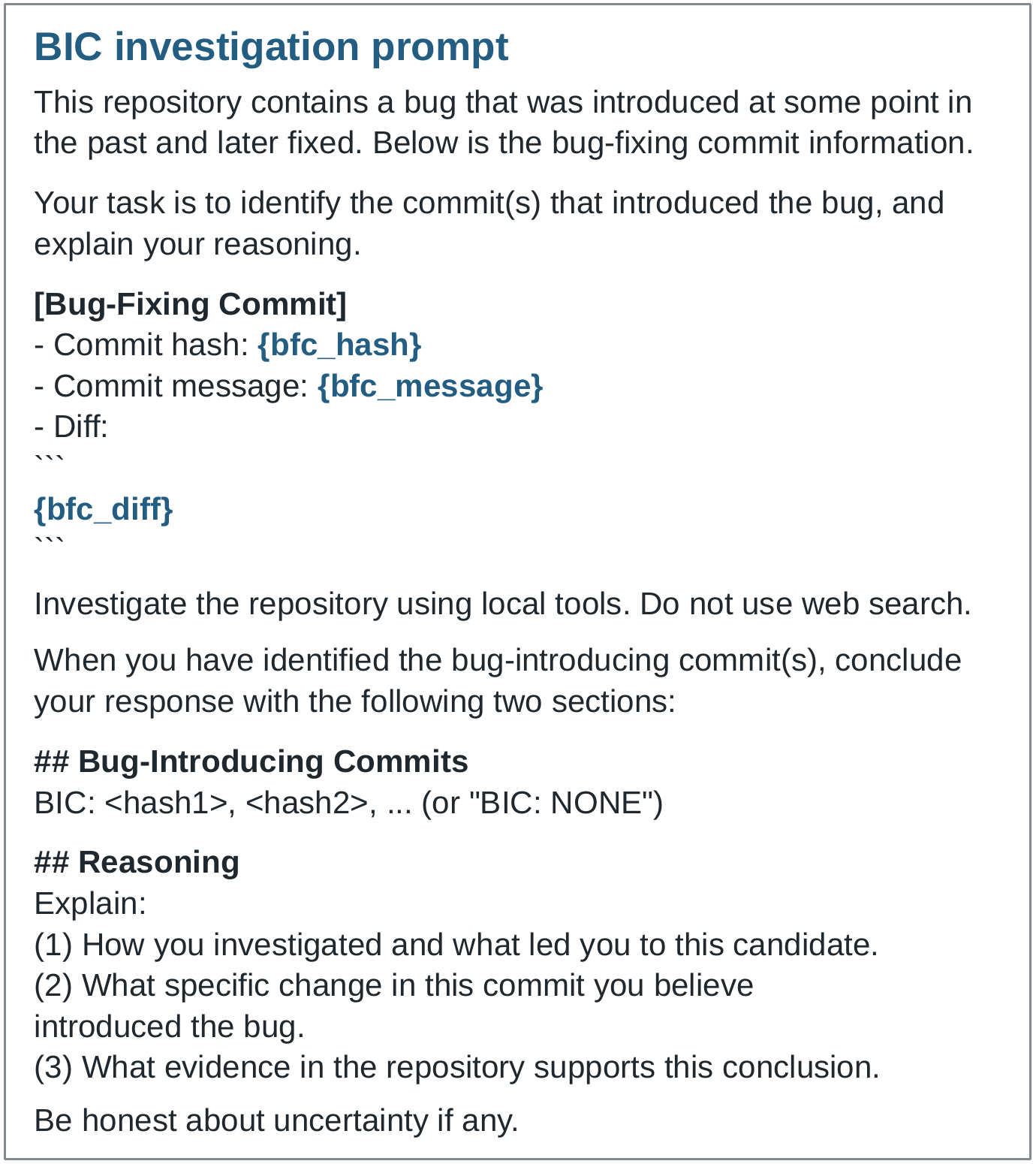}
  \caption{BIC investigation prompt with case-specific placeholders.}
  \label{fig:investigation-prompt}
\end{figure}

We use the prepared repository as the working directory and pass the prompt in Figure~\ref{fig:investigation-prompt} to \tool{}.
For each case, we insert the recreated BFC's hash, edited message, and diff.
The prompt asks \tool{} to identify the BICs and explain its reasoning.
It does not specify Git commands, candidates, or a search order.
We request one or more commit hashes, an account of the investigation, the bug-introducing change, and supporting evidence.
If \tool{} finds no applicable commit, it returns \texttt{BIC: NONE}.

\subsection{Model Communication Audit}
\label{subsec:execution-records}

We audit model communication to check for unintended answer leakage.
A proxy records the inputs sent to the model and the responses returned to \tool{}.
For each investigation, we also save the final response, tool calls and their results, timestamps, and resource usage.
Authentication headers are excluded from the captures.
We check the correspondence between requests and responses and verify that responses are recorded through completion.
We trace where each oracle BIC hash first appears in the initial Git context, tool results, and model responses.
A hash in an ordinary commit listing does not by itself reveal which commit introduced the bug.

\section{Study Design}
\label{sec:study-design}

This section presents our research questions, study workflow, and evaluation dataset.
The following sections describe the methods and results for each RQ.

\subsection{Overview}
\label{subsec:study-overview}

We evaluate \tool{} from three perspectives: identification accuracy, search strategies, and failure causes.

\noindent\textbf{RQ1: \rqone{}}
We compare \tool{}'s identification accuracy with that of existing rule-based SZZ methods and LLM- and agent-based SZZ methods.
We also compare the execution time and cost required to identify BICs.

\noindent\textbf{RQ2: \rqtwo{}}
We examine how \tool{} investigates BICs without being instructed to start with code tracing.
Using execution logs, we analyze he number of tool calls, command usage, and the order of investigation operations.
This analysis characterizes the search trajectories \tool{} follows to identify BICs.

\noindent\textbf{RQ3: \rqthree{}}
We analyze cases in which \tool{} failed to identify BICs correctly.
To investigate the causes of failure in detail, we manually examine the changes and histories of the BFC, the oracle BICs, and the commits selected by \tool{}, alongside its explanations and execution logs.

\begin{figure}[!t]
  \centering
  \includegraphics[width=\columnwidth]{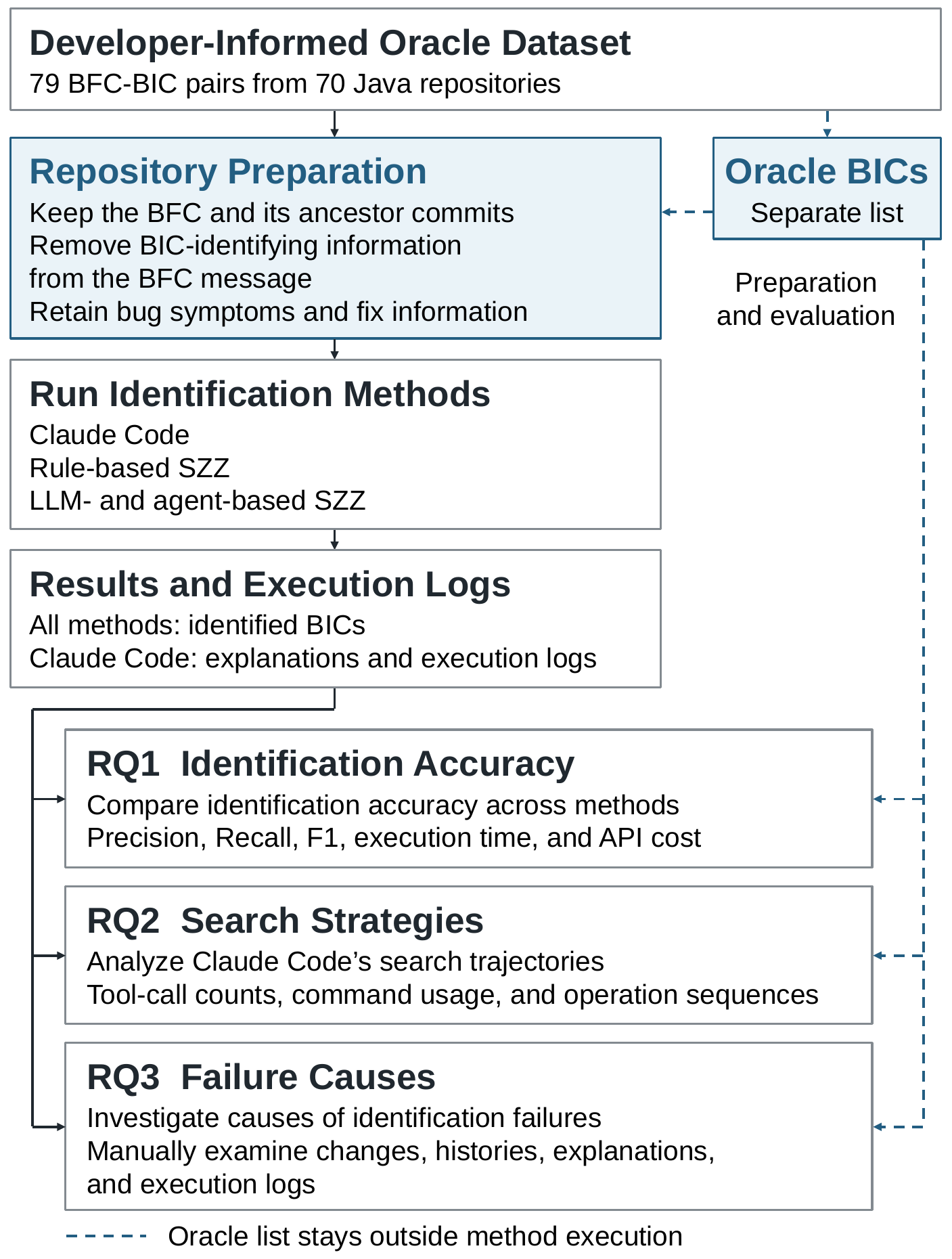}
  \caption{Overview of this study.}
  \label{fig:study-overview}
\end{figure}

Figure~\ref{fig:study-overview} shows the study workflow.
We use each method's identification results for the comparison in RQ1.
For \tool{}, we analyze its search process in RQ2 and its identification failures in RQ3.

\subsection{Evaluation Dataset}
\label{subsec:dataset}

We use the Developer-Informed Oracle Dataset constructed by Rosa et al.~\cite{rosa2023comprehensive}.
It contains cases in which developers explicitly identified BICs in BFC messages.
Rosa et al. manually reviewed the extracted messages to confirm that they clearly identified the BICs.
The resulting labels therefore draw on developer knowledge rather than automated SZZ results.
Rosa et al.'s developer-informed data has also been used to evaluate LLM4SZZ, AgentSZZ, SZZ-Agent, and Simple-SZZ-Agent~\cite{tang2025llm4szz,lyu2026agentszz,risse2026agents}.

We focus on Java, which is evaluated in these prior studies.
This language restriction also keeps API costs and manual review manageable across multiple LLM- and agent-based methods.
From version~1 of the official replication package,\footnote{\url{https://doi.org/10.6084/m9.figshare.19586500.v1}}
we select all 81 cases whose BFC-modified files are labeled Java in the dataset.
We exclude the case from \texttt{Kasukoi/eisverkauf} because the repository was unavailable, leaving 80 eligible cases.
A Liferay case is also excluded from execution after three attempts to clone its full history failed.\footnote{The dataset lists \texttt{brianchandotcom/liferay-portal}; we attempted to acquire its history from \texttt{liferay/liferay-portal}.}
The final evaluation dataset contains 79 BFC--BIC pairs from 70 repositories.

\section{Identification Accuracy}
\label{sec:rq1}
\subsection{Approach}
\label{subsec:rq1-approach}

We compare \tool{} with rule-based SZZ methods and LLM- and agent-based SZZ methods in terms of identification accuracy and cost.

\subsubsection{Baselines}

Our rule-based baselines are B-SZZ, AG-SZZ, MA-SZZ, R-SZZ, L-SZZ, and RA-SZZ.
These six variant families were evaluated in both the LLM4SZZ and SZZ-Agent studies~\cite{tang2025llm4szz,risse2026agents}.
We use the implementations in PySZZ v2,\footnote{\url{https://github.com/grosa1/pyszz_v2}} provided by Rosa et al.~\cite{rosa2023comprehensive}.

Our LLM- and agent-based SZZ baselines are LLM4SZZ, AgentSZZ, SZZ-Agent, and Simple-SZZ-Agent~\cite{tang2025llm4szz,lyu2026agentszz,risse2026agents}.
We use the replication packages of LLM4SZZ (Figshare v1) and AgentSZZ, adapting their model interfaces to access Claude through an API.
We use the replication packages of SZZ-Agent and Simple-SZZ-Agent, changing their oracle-based early-exit check.
The original packages stop the investigation if the oracle BIC is absent from the prepared search space.
Because the oracle BIC is unknown in practical use, we remove this check and allow both methods to continue their investigations and produce an answer.
We also considered AgenticSZZ~\cite{shi2026agenticszz}, but discontinued its evaluation for cost reasons after investigations in two of the first three cases hit resource limits under a USD~5 per-run budget.
AgenticSZZ identified the correct BIC in four of its five completed runs, while \tool{} was correct in all five corresponding runs.

\subsubsection{Execution Conditions}

We use the same 79 cases for all methods, with the BFC messages prepared as described in Section~\ref{sec:approach}.
We evaluate each rule-based method once per case and each LLM- or agent-based method in three independent runs.
All LLM- and agent-based methods use Sonnet 5.
\tool{}, SZZ-Agent, and Simple-SZZ-Agent use Claude Code 2.1.232 with reasoning effort set to \texttt{high}.
LLM4SZZ and AgentSZZ use API interfaces without an explicit reasoning-effort setting.

We limit each Claude Code invocation to 600 seconds.
For SZZ-Agent, we also impose a 3,600-second limit across its two stages.
LLM4SZZ and AgentSZZ each have a 600-second limit for the entire method execution.
Each LLM- or agent-based attempt has a USD~5 stopping threshold, based on reported cost for Claude Code and estimated cost for API-based methods.

\subsubsection{Evaluation Metrics}

We count correctly identified BICs as true positives (TP), incorrectly selected commits as false positives (FP), and missed oracle BICs as false negatives (FN).
We score all 79 cases in each repetition, including unresolved execution failures.
Runs that return no candidates or remain incomplete are scored as having identified no BICs, with all oracle BICs counted as FN.
For each repetition, we sum TP, FP, and FN across all cases and calculate Precision, Recall, and F1.
For methods with three repetitions, we report the mean TP, FP, Precision, Recall, and F1 across the repetitions.
Precision measures the proportion of selected commits that are correct, while Recall measures the proportion of oracle BICs recovered.
F1 score is their harmonic mean~\cite{manning2008introduction}.
These metrics are used in prior SZZ evaluations~\cite{rosa2023comprehensive,tang2023neural,lyu2024evaluating,perezrosero2024wia,tang2025llm4szz,lyu2026agentszz,risse2026agents}.

For statistical comparison, we calculate each case's mean F1 over all repetitions, assigning zero to incomplete runs.
We compare the resulting case-level scores of \tool{} and each baseline using a two-sided Wilcoxon signed-rank test~\cite{Wilcoxon1945Biometrics}.
All ten comparisons use the same 79 paired cases.
We apply Holm's correction across the ten comparisons~\cite{holm1979simple} and report adjusted $p$-values at significance levels of 0.05 and 0.01.

We report mean elapsed time and monetary cost over normally completed runs, consistent with prior SZZ studies~\cite{tang2025llm4szz,lyu2026agentszz,risse2026agents}.
Elapsed time covers the entire method execution, including preparation and communication, but excludes quota-related waiting and discarded attempts.
We calculate model API cost for all LLM- and agent-based methods from input, output, and cache token counts using the same token prices.

\begin{table}[!t]
	\centering
	\caption{Accuracy and cost comparison between Claude Code and existing SZZ methods.}
	\label{tab:accuracy-results}
	\fontsize{8}{9.5}\selectfont
	\setlength{\tabcolsep}{2pt}
	\renewcommand{\arraystretch}{1.15}
	\begin{tabular*}{\columnwidth}{@{\extracolsep{\fill}}lrrrrrrr@{}}
		\toprule
		Method & TP\hphantom{.00} & FP\hphantom{.00} & Pre. & Rec. & F1\makebox[0.8em][l]{} & sec. & \$ \\
		\midrule
		\multicolumn{8}{@{}l}{\textbf{Rule-based SZZ}} \\
		B-SZZ & 52\hphantom{.00} & 134\hphantom{.00} & 0.280 & 0.658 & 0.392\makebox[0.8em][l]{$^{**}$} & -- & 0.000 \\
		AG-SZZ & 47\hphantom{.00} & 133\hphantom{.00} & 0.261 & 0.595 & 0.363\makebox[0.8em][l]{$^{**}$} & -- & 0.000 \\
		MA-SZZ & 45\hphantom{.00} & 141\hphantom{.00} & 0.242 & 0.570 & 0.340\makebox[0.8em][l]{$^{**}$} & -- & 0.000 \\
		R-SZZ & 37\hphantom{.00} & 32\hphantom{.00} & 0.536 & 0.468 & 0.500\makebox[0.8em][l]{$^{**}$} & -- & 0.000 \\
		L-SZZ & 33\hphantom{.00} & 36\hphantom{.00} & 0.478 & 0.418 & 0.446\makebox[0.8em][l]{$^{**}$} & -- & 0.000 \\
		RA-SZZ & 39\hphantom{.00} & 119\hphantom{.00} & 0.247 & 0.494 & 0.329\makebox[0.8em][l]{$^{**}$} & -- & 0.000 \\
		\midrule
		\multicolumn{8}{@{}l}{\textbf{LLM- and agent-based SZZ}} \\
		LLM4SZZ & 46.67 & 28.33 & 0.622 & 0.591 & 0.606\makebox[0.8em][l]{$^{**}$} & 51.91 & \textbf{0.090} \\
		AgentSZZ & 50.67 & \textbf{18.33} & 0.735 & 0.641 & 0.685\makebox[0.8em][l]{$^{**}$} & \textbf{33.32} & 0.137 \\
		SZZ-Agent & 48.67 & 25.67 & 0.655 & 0.616 & 0.635\makebox[0.8em][l]{$^{**}$} & 112.82 & 0.453 \\
		Simple-SZZ-Agent & 55.00 & 24.00 & 0.696 & 0.696 & 0.696\makebox[0.8em][l]{$^{*}$} & 80.99 & 0.266 \\
		\midrule
		\textbf{Claude Code} & \textbf{61.33} & 18.67 & \textbf{0.767} & \textbf{0.776} & \textbf{0.771}\makebox[0.8em][l]{} & 50.14 & 0.166 \\
		\bottomrule
	\end{tabular*}
	\par\smallskip
	\begin{minipage}{\columnwidth}
		\footnotesize Bold indicates the best values in each column (time and cost: LLM- and agent-based methods only).
		Asterisks indicate significant F1 differences from Claude Code (Holm-adjusted $^{*}p<0.05$, $^{**}p<0.01$).
	\end{minipage}
\end{table}

\subsection{Results}

\textbf{\tool{} outperformed all baselines in Precision, Recall, and F1 score.}
Table~\ref{tab:accuracy-results} shows that \tool{} achieved a Precision of 0.767, a Recall of 0.776, and an F1 score of 0.771.
Simple-SZZ-Agent was the strongest baseline, with an F1 score of 0.696, followed by AgentSZZ at 0.685.
Without prepared candidate histories or SZZ-specific investigation instructions, \tool{} surpassed these baselines by 0.075 and 0.087 F1 points, respectively.

\textbf{\tool{}'s F1 scores were significantly higher than those of all ten baselines.}
The comparison between \tool{} and Simple-SZZ-Agent gave a Holm-adjusted $p=0.0362$.
Each comparison between \tool{} and the other nine baselines gave an adjusted $p<0.01$.

\textbf{\tool{} had a mean API cost of USD~0.166 and a mean running time of 50.14 seconds per run.}
Among the LLM- and agent-based methods, LLM4SZZ had the lowest API cost at USD~0.090.
\tool{} cost about 1.85 times as much.
AgentSZZ was the fastest at 33.32 seconds, with \tool{} taking about 1.50 times as long.
Compared with Simple-SZZ-Agent, the strongest baseline in F1 score, \tool{} reduced API cost by 37.7\% and running time by 38.1\%.
It achieved these reductions without SZZ-specific investigation instructions, specialized tools, or a predefined candidate search space.

\summarybox{Answer to RQ1}{%
	\tool{} outperformed all evaluated baselines in Precision (0.767), Recall (0.776), and F1 score (0.771).
	\tool{}'s mean API cost was USD~0.166 per run, and its mean running time was 50.14 seconds per run.
	These results suggest a good balance between accuracy, running time, and API cost.
}

\section{Search Strategies}
\label{sec:rq2}
In RQ2, we examine how \tool{} searches for BICs without being instructed to start with code tracing.
We analyze the commands it uses, the information it requests at the start, and the order of its investigation operations.
We measure investigation effort using the number of Bash calls.

\subsection{Approach}
\label{subsec:rq2-approach}

\subsubsection{Tool and command usage}

We extract Bash calls and the commands they contain from execution logs of three runs on each of the 79 cases.
We count each command within a Bash call separately; for example, \texttt{git log -{}-oneline | head -20} counts as one Bash call and two commands.
We group commands into Git and non-Git commands and count their occurrences to identify the operations \tool{} uses most often.
To examine what information \tool{} requests at the start of its search, we inspect the first explicit Git command and its arguments in each run.

\begin{table}[!t]
	\centering
	\caption{Investigation operation categories.}
	\label{tab:rq2-operation-categories}
	\small
	\setlength{\tabcolsep}{4pt}
	\renewcommand{\arraystretch}{1.15}
	\begin{tabularx}{\linewidth}{@{}l l >{\raggedright\arraybackslash}X@{}}
		\toprule
		Label & Operation             & Command examples                                                                                                                                        \\
		\midrule
		A     & History search        & \texttt{git log -{}- path};\newline \texttt{git blame path}                                                                                             \\
		\addlinespace[3pt]
		B     & Commit inspection     & \texttt{git show REV};\newline \texttt{git diff REV1 REV2};\newline \texttt{git log -1 REV}                                                             \\
		\addlinespace[3pt]
		C     & Code search / reading & \texttt{git grep PATTERN};\newline \texttt{git show REV:path};\newline file-targeted \texttt{rg}, \texttt{grep},\newline \texttt{cat}, and \texttt{sed} \\
		\bottomrule
	\end{tabularx}
	\par\smallskip
	\begin{minipage}{\linewidth}
		\footnotesize
		\texttt{REV} denotes a specified commit.
		Categories depend on arguments: single-commit \texttt{log} is B only without path or search filters; output filters retain the source operation's category.
	\end{minipage}
\end{table}

\subsubsection{Investigation process}
We analyze operation sequences to examine how \tool{} begins its investigations and whether it follows different paths to the same answer.
We developed a script that classifies investigative operations into history search (A), commit inspection (B), and code search or reading (C), based on command names and arguments.
Table~\ref{tab:rq2-operation-categories} lists these categories and command examples.
The script classified 1,592 of the 1,612 Bash calls.
We manually classified the remaining 20 calls from their commands and outputs.

We construct operation sequences from these labels and merge consecutive repetitions of the same label; for example, AABBA becomes ABA.
We analyze operation patterns only within segments where the order can be determined from the logs.
We do not join operations across segments with uncertain order.
We also compare operation sequences within each case where all three runs return the same BIC set.

\subsubsection{Investigation effort}
We count Bash calls per run to measure the investigation effort.
We classify a run as successful if its returned BIC set exactly matches the oracle, and as unsuccessful otherwise.
We compare the distributions, medians, and means of Bash calls between the two groups.
Since a Bash call can contain multiple commands, we also calculate the mean number of commands per Bash call.

\subsection{Results}
\label{subsec:rq2-results}

\begin{table}[!t]
	\centering
	\caption{Git and non-Git command usage across 237 investigations.}
	\label{tab:rq2-command-usage}
	\begingroup
\footnotesize
\setlength{\tabcolsep}{3pt}
\renewcommand{\arraystretch}{1.12}
\begin{tabular*}{\linewidth}{@{\extracolsep{\fill}}lrlr@{}}
\toprule
Git command & Count (\%) & Non-Git command & Count (\%) \\
\midrule
\texttt{show} & 883 (48.8) & \texttt{head} & 558 (28.8) \\
\texttt{log} & 816 (45.1) & \texttt{grep} & 547 (28.2) \\
\texttt{merge-base} & 35 (1.9) & \texttt{echo} & 345 (17.8) \\
\texttt{blame} & 24 (1.3) & \texttt{sed} & 204 (10.5) \\
\texttt{grep} & 18 (1.0) & \texttt{tail} & 93 (4.8) \\
\texttt{cat-file} & 15 (0.8) & \texttt{wc} & 51 (2.6) \\
Other (7 types) & 17 (0.9) & Other (15 types) & 142 (7.3) \\
\midrule
\textbf{Total} & \textbf{1,808 (100.0)} & \textbf{Total} & \textbf{1,940 (100.0)} \\
\midrule
\multicolumn{3}{@{}l}{All commands (34 types)} & 3,748 \\
\bottomrule
\end{tabular*}
\endgroup

\end{table}

\begin{table}[!t]
	\centering
	\caption{Commands and output formats of the first explicit Git request in each run.}
	\label{tab:rq2-first-git}
	\begingroup
\small
\renewcommand{\arraystretch}{1.12}
\begin{tabular*}{\linewidth}{@{\extracolsep{\fill}}lrr@{}}
\toprule
First explicit Git request & Runs & \% \\
\midrule
\textbf{\texttt{log}} & 234 & 98.7 \\
\quad\texttt{-{}-oneline}, no patch & 141 & 59.5 \\
\qquad List request & 129 & 54.4 \\
\qquad One-commit request (\texttt{-1}) & 12 & 5.1 \\
\quad\texttt{-{}-oneline}, with patch & 17 & 7.2 \\
\quad Standard format, with patch & 76 & 32.1 \\
\texttt{show} & 3 & 1.3 \\
\midrule
\textbf{Total} & \textbf{237} & \textbf{100.0} \\
\bottomrule
\end{tabular*}
\endgroup

	\par\smallskip
\end{table}

\textbf{\tool{} mainly used \texttt{git log} and \texttt{git show} (94.0\% of Git command occurrences), while rarely using \texttt{git blame} (1.3\%), which is widely used in traditional SZZ methods.}
As shown in Table~\ref{tab:rq2-command-usage}, \texttt{git show} appeared 883 times and \texttt{git log} 816 times, compared with only 24 occurrences of \texttt{git blame}.
\texttt{git show} displays the changes made by a specified commit, whereas \texttt{git log} lists the commit history.
These usage patterns indicate that \tool{} mainly searched commit histories and examined individual changes in detail.
Among non-Git commands, \texttt{grep}, \texttt{head}, and \texttt{sed} were frequently used.
These commands were used to search and read code or filter Git output.
\begin{figure}[!t]
  \centering
  \includegraphics[width=\linewidth]{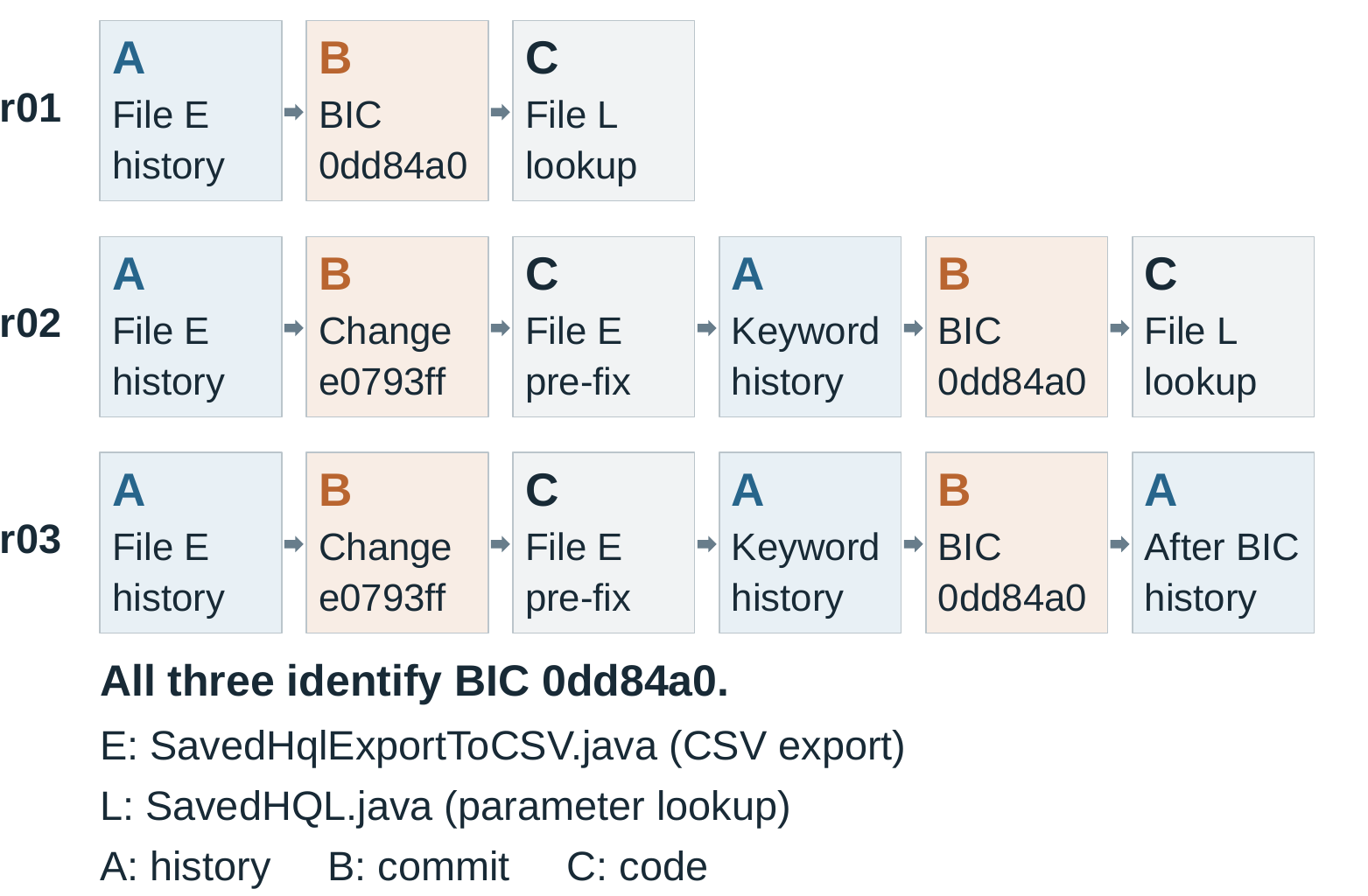}
  \caption{Three investigation paths to the same BIC in case ID 3109.}
  \label{fig:rq2-case}
\end{figure}

\textbf{\tool{} began more than half of its investigations by listing commit-message subjects.}
As shown in Table~\ref{tab:rq2-first-git}, the first explicit Git request was \texttt{git log} in 234 runs and \texttt{git show} in the remaining three.
In 129 runs (54.4\%), \tool{} first requested a \texttt{git log -{}-oneline} listing of commit hashes and message subjects without diffs.
This suggests that \tool{} used semantic information from commit messages in its search.

\textbf{\tool{} began 189 of the 237 runs (79.7\%) with history search (A) followed by commit inspection (B).}
Code tracing finds candidates by tracing the lines changed by a fix back through history.
In contrast, we observed \tool{} reviewing the histories of files changed by the BFC and using commit messages and diffs to find candidates for closer inspection.
\tool{} also used code search and reading (C) to examine surrounding code and called methods, extending its search to other files or investigating how a change affected program behavior.

\textbf{\tool{} followed different investigation paths in 62 of the 68 cases (91.2\%) where all three runs returned the same BIC set.}
The returned BIC set matched the oracle in 57 of these cases and differed from it in the remaining 11.
As shown in Figure~\ref{fig:rq2-case}, all three runs in case 3109 identified the same oracle BIC, \texttt{0dd84a0}, but their operation sequences were ABC, ABCABC, and ABCABA.
In the first run, \tool{} began commit inspection with the oracle BIC, then examined a called method's implementation and ended the investigation.
In contrast, the second and third runs first inspected a different commit.
\tool{} then read the pre-fix code, returned to history search, and inspected the oracle BIC's diff.

\begin{figure}[!t]
	\centering
	\includegraphics[width=\linewidth]{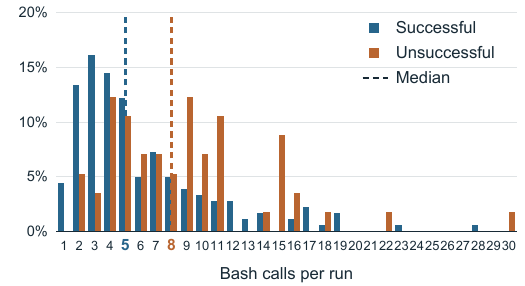}
	\caption{Distribution of Bash calls in 180 successful and 57 unsuccessful runs. }
	\label{fig:rq2-bash-distribution}
\end{figure}

\textbf{\tool{} made a median of 5 Bash calls in successful runs and 8 in unsuccessful runs, as shown in Figure~\ref{fig:rq2-bash-distribution}.}
The corresponding means were 6.14 and 8.88 calls.
Each Bash call contained an average of 2.33 commands.

\FloatBarrier
\summarybox{Answer to RQ2}{%
	\tool{} typically listed commit histories and then inspected individual commits, rather than starting with code tracing.
	It frequently examined commit-message subjects, suggesting that it used semantic information in its search.
	Even when all three runs returned the same BIC set, \tool{} followed different investigation paths in 62 of the 68 cases.
}

\section{Failure Causes}
\label{sec:rq3}
\subsection{Approach}
\label{subsec:rq3-approach}

To investigate the causes of identification failure, we examined the 22 cases in which the BIC set identified by \tool{} did not match the oracle in at least one of the three runs.
For each case, we compared the diffs and histories of the BFC, the oracle BICs, and the commits identified by \tool{}.
We also reviewed \tool{}'s explanations and execution logs.
We assessed whether the oracle BICs and the commits identified by \tool{} introduced the bug addressed by the BFC.

Two authors analyzed each case using open coding and grouped the resulting codes into a classification scheme.
The authors initially disagreed on two cases.
They reviewed the evidence and discussed these cases until they agreed on the final classification scheme and case assignments.

We assessed whether the oracle BIC and \tool{}'s answer were valid for the bug addressed by the BFC.
If the oracle was valid but \tool{}'s answer was incorrect, we classified the case as \textit{\tool{} was incorrect}.
If the oracle was invalid, we classified the case as \textit{The oracle was incorrect}.
If both the oracle and \tool{}'s answer had supporting evidence, we classified the case as \textit{Ambiguous or incomplete BIC}.
These cases involved different views of when a bug was introduced or what counts as a bug, equivalent changes on different branches, or multiple bug fixes in the BFC.

\subsection{Results}
\label{subsec:rq3-results}

\textbf{Surprisingly, disagreement with the oracle did not always mean that \tool{} was wrong.}
We investigated why \tool{} disagreed with the oracle.
The 22 cases were classified into three categories, as summarized in Table~\ref{tab:rq3-disagreement-causes}.

\begin{table}[!t]
	\centering
	\caption{Reasons for disagreement with the oracle.}
	\label{tab:rq3-disagreement-causes}
	\footnotesize
	\setlength{\tabcolsep}{3pt}
	\renewcommand{\arraystretch}{1.15}
	\begin{tabularx}{\linewidth}{@{}>{\raggedright\arraybackslash}Xc>{\raggedright\arraybackslash}p{0.33\linewidth}@{}}
		\toprule
		Category / reason                                                 & Cases      & Case IDs                                \\
		\midrule
		\textbf{\tool{} was incorrect}                                    & \textbf{6} & 921, 1065, 1666, 1991, 2602, 2854       \\
		\midrule
		\textbf{Oracle was incorrect}                                     & \textbf{7} & 907, 1121, 1381, 1728, 2557, 2911, 3530 \\
		\midrule
		\textbf{Ambiguous or incomplete BIC}                              & \textbf{9} &                                         \\
		\addlinespace[3pt]
		\hangindent=0.75em\hangafter=0 Introducing vs. triggering the bug & (4)        & 313, 399, 717, 2513                     \\
		\addlinespace[3pt]
		\hangindent=0.75em\hangafter=0 Disagreement about what a bug is   & (2)        & 2357, 3521                              \\
		\addlinespace[3pt]
		\hangindent=0.75em\hangafter=0 Equivalent commits across branches & (2)        & 975, 1470                               \\
		\addlinespace[3pt]
		\hangindent=0.75em\hangafter=0 Multiple bug fixes in the BFC      & (1)        & 718                                     \\
		\midrule
		Total                                                             & 22         &                                         \\
		\bottomrule
	\end{tabularx}
	\par\smallskip
\end{table}

\noindent
\textbf{\tool{} was incorrect:}
In four of the six cases, \tool{} selected a related commit instead of the one that introduced the bug.
For example, in case 2854, it selected an incomplete fix rather than the oracle commit that had removed a required class-loader setting.
In another case, it failed to investigate a relevant configuration change.
In the remaining case, it correctly identified the cause of the bug but attributed the responsible change to the wrong commit.

\noindent
\textbf{The oracle was incorrect:}
In seven cases, the oracle identified the wrong commit.
For example, in case 1381, according to the oracle, the BFC fixed a bug in the Android code, but the oracle's BIC incorrectly pointed to a related change that affected only iOS code and not the Android code (probably a user error, as the true BIC was the parent of the oracle's BIC).
Supporting evidence for all seven cases is included in our replication package.

\noindent
\textbf{Ambiguous or incomplete BIC:}
In four cases, an earlier commit introduced code with a latent bug, but a later commit triggered it by changing how the code was used.
\tool{} selected the earlier commit while the oracle identified the later one, but the reverse also occurred.
To verify that the later commit triggered the bug, we tested the affected code before and after the commit.
In all four cases, the test passed with the same input before the commit, failed after it, and passed again after the BFC.
Two cases concerned whether static-analysis warnings and performance regressions count as bugs.
In some runs, \tool{} did not regard these issues as functional bugs and returned \texttt{NONE}.
In two cases, the same identical bug-inducing change appeared under a different commit ids on different branches , and \tool{} selected the corresponding commit instead of the oracle BIC.
In the remaining case, the BFC fixed two bugs, but the oracle recorded only one and \tool{} identified the
other.

\FloatBarrier
\summarybox{Answer to RQ3}{%
	Manual inspection of the 22 cases showed that disagreements with the oracle were not always due to \tool{} errors.
	Six cases involved \tool{} errors, seven involved incorrect oracle labels, and nine involved ambiguous or incomplete BIC labels.
}

\section{The Importance of Commit Messages}
\label{sec:commit-messages}

In RQ2, \tool{} used semantic information in commit messages to identify BICs rather than starting with code tracing.
Because commit messages describe what was changed and why, their availability may affect BIC identification accuracy and search efficiency.
We therefore removed commit messages and examined how identification accuracy, the number of Bash calls, execution time, and cost changed.

We removed all commit messages from the repositories while preserving the source code and parent--child relationships between commits.
We also removed the BFC message from the initial prompt.
We evaluated the same 79 cases with three runs per case under the execution conditions described in Section~\ref{sec:rq1}.
As in RQ1, we report TP, FP, Precision, Recall, and F1 averaged over the three repetitions.

\textbf{\tool{} could still identify BICs without commit messages, but made more Bash calls, took longer, and cost more.}
As shown in Table~\ref{tab:commit-message-ablation}, the mean number of Bash calls per run was 1.33 times as high without messages as with them.
Without messages, \tool{} needed additional commands to understand changes from code and diffs.
Mean execution time and API-equivalent cost per run were also 1.26 and 1.20 times as high, respectively.
The increases in all three metrics were statistically significant after Holm correction.
In contrast, the F1 score decreased from 0.771 to 0.751, but the difference was not statistically significant in the case-level comparison.
These results suggest that commit messages help narrow down the changes to investigate and reduce the effort needed to identify BICs.

\begin{table}[!t]
	\centering
	\caption{Accuracy and search effort with and without commit messages.}
	\label{tab:commit-message-ablation}
	\fontsize{8}{9.5}\selectfont
	\setlength{\tabcolsep}{1.5pt}
	\renewcommand{\arraystretch}{1.15}
	\begin{tabular*}{\columnwidth}{@{\extracolsep{\fill}}lrrrrrrrr@{}}
		\toprule
		\shortstack[l]{Commit\\messages} & TP & FP & Pre. & Rec. & F1 & \shortstack[r]{Bash\\calls}\makebox[0.8em][l]{} & \shortstack[r]{Time\\(s)}\makebox[0.8em][l]{} & \shortstack[r]{Cost\\(USD)}\makebox[0.8em][l]{} \\
		\midrule
		With & 61.33 & 18.67 & 0.767 & 0.776 & 0.771 & 6.80\makebox[0.8em][l]{} & 48.61\makebox[0.8em][l]{} & 0.166\makebox[0.8em][l]{} \\
		\textbf{Without} & 59.33 & 19.67 & 0.751 & 0.751 & 0.751 & 9.01\makebox[0.8em][l]{$^{**}$} & 61.21\makebox[0.8em][l]{$^{*}$} & 0.198\makebox[0.8em][l]{$^{**}$} \\
		\bottomrule
	\end{tabular*}
	\par\smallskip
	\raggedright
	Holm-adjusted $^{*}p<0.05$, $^{**}p<0.01$.
\end{table}

\summarybox{Implications}{
	Using commit messages to narrow down candidate commits before inspecting diffs is a promising way to reduce investigation effort in SZZ methods.
}

\section{Improving the Dataset}
\label{sec:dataset-reconstruction}

RQ3 revealed incorrect oracle labels and cases with ambiguous BIC definitions in the Developer-Informed Oracle Dataset.
These issues make it difficult to evaluate the answers returned by SZZ methods accurately.
We therefore need to address these label errors and ambiguities to evaluate each method's BIC identification accuracy reliably.

\subsection{Approach}
\label{subsec:approach}

We created \textit{Developer-Informed Oracle Verified} (DIO-V) by reconstructing the Developer-Informed Oracle Dataset to evaluate SZZ methods reliably.
To prevent ambiguity in BIC definitions and labels from affecting the evaluation, we focused on cases for which we could establish a single oracle BIC.
We therefore excluded the nine cases classified as \textit{Ambiguous or incomplete BIC} in RQ3.

We corrected the oracle labels for four of the seven cases with label errors identified in RQ3.
These corrections were based on commit diffs and reproduction tests where available.
We excluded the remaining three cases because the public history was insufficient to establish the BIC or multiple changes played different causal roles, preventing us from defining a single oracle BIC.
We retained the original labels for all other cases, leaving 67 cases in DIO-V.
We provide DIO-V in our replication package.

We use DIO-V to reevaluate the identification accuracy of \tool{} and the four LLM- and agent-based SZZ baselines.
We rescore the saved answers from the three runs per case in RQ1 using the DIO-V oracle labels, without rerunning the methods.
As in RQ1, we report TP, FP, Precision, Recall, and F1 averaged over the three repetitions.
We compare the mean F1 scores for each case using a two-sided Wilcoxon signed-rank test.
We apply Holm's correction across the four comparisons between \tool{} and the baselines.

Accuracy scores do not show whether methods identify the same cases or different ones.
We therefore compare the overlap between the cases identified by \tool{} and each baseline.
For all methods, we consider a case identified if at least one of the three final answers contains the DIO-V oracle BIC.

\subsection{Results}
\label{subsec:reevaluation}

\begin{table}[!t]
	\centering
	\caption{BIC identification accuracy on the DIO-V (67 cases).}
	\label{tab:revised-oracle-results}
	\footnotesize
	\setlength{\tabcolsep}{2pt}
	\renewcommand{\arraystretch}{1.15}
	\begin{tabular*}{\columnwidth}{@{\extracolsep{\fill}}lrrrrr@{}}
		\toprule
		Method & TP & FP & Precision & Recall & F1 score\makebox[0.8em][l]{} \\
		\midrule
		LLM4SZZ & 47.33 & 15.67 & 0.751 & 0.706 & 0.728\makebox[0.8em][l]{$^{**}$} \\
		AgentSZZ & 51.33 & 9.00 & 0.851 & 0.766 & 0.806\makebox[0.8em][l]{$^{**}$} \\
		SZZ-Agent & 50.00 & 15.00 & 0.769 & 0.746 & 0.758\makebox[0.8em][l]{$^{**}$} \\
		Simple-SZZ-Agent & 55.33 & 11.67 & 0.826 & 0.826 & 0.826\makebox[0.8em][l]{$^{**}$} \\
		\midrule
		\textbf{\tool{}} & \textbf{63.00} & \textbf{4.33} & \textbf{0.936} & \textbf{0.940} & \textbf{0.938}\makebox[0.8em][l]{} \\
		\bottomrule
	\end{tabular*}
	\par\smallskip
	\raggedright
	Holm-adjusted $^{**}p<0.01$ versus \tool{} (F1).
\end{table}

\textbf{\tool{} achieved the highest identification accuracy on DIO-V.}
Table~\ref{tab:revised-oracle-results} shows that \tool{} achieved a Precision of 0.936, a Recall of 0.940, and an F1 score of 0.938.
The strongest baseline, Simple-SZZ-Agent, achieved an F1 score of 0.826.
\tool{}'s F1 scores were significantly higher than those of all four baselines after Holm correction ($p<0.01$).
For three of the four cases with corrected oracle labels, \tool{} returned the corrected BIC in all three runs.
In the remaining case, two of the three runs returned the corrected BIC.
These results show that incorrect oracle labels led us to underestimate \tool{}'s ability to identify BICs in these cases.

\begin{figure}[!t]
	\centering
	\includegraphics[width=\columnwidth]{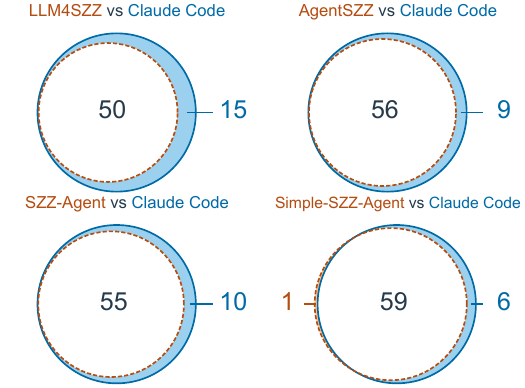}
	\caption{Overlap of cases correctly identified by \tool{} and each baseline on DIO-V.}
	\label{fig:pairwise-case-overlap}
\end{figure}

\textbf{\tool{} identified all but one of the cases identified by the baselines.}
Figure~\ref{fig:pairwise-case-overlap} shows that \tool{} identified the oracle BIC in 65 of the 67 cases in at least one of the three runs.
These included every case identified by LLM4SZZ, AgentSZZ, or SZZ-Agent, and 59 of the 60 cases identified by Simple-SZZ-Agent.
The only exception was case 921.
None of the five methods identified the oracle BIC in case 2602.

In case 921, \tool{} did not sufficiently investigate the relevant changes and missed the oracle BIC in all three runs.
\tool{} selected the commit that introduced the code changed by the fix, without examining a later change that removed timezone settings.
Simple-SZZ-Agent identified the oracle BIC in only one of its three runs, so it did not consistently identify this case either.

\summarybox{Implications}{%
	\tool{} achieved high identification accuracy and identified the correct BICs in the relabeled cases.
	These findings highlight the importance of reviewing and correcting ground-truth labels in SZZ research, rather than assuming that existing labels are always correct.
}

\FloatBarrier
\section{Anti-Patterns in SZZ Search Design}
\label{sec:performance-differences}

\tool{} achieved higher BIC identification accuracy than the four baselines, even though all five methods used the same model, Sonnet 5.
Unlike \tool{}, these baselines use SZZ-specific search instructions or restricted search spaces designed by researchers.
To develop SZZ methods with higher accuracy and lower cost, we need to learn from anti-patterns in existing search designs.
We manually analyze the methods' designs and execution logs for cases in which \tool{} identified the BIC but a baseline did not.

\textbf{Researcher-defined search spaces sometimes excluded the oracle BIC.}
LLM4SZZ uses \texttt{git blame} to generate candidate commits and then examines them to select a BIC.
In 11 of the 15 cases identified by \tool{} but not LLM4SZZ, the oracle BIC was absent from these candidates in all three runs.
Simple-SZZ-Agent restricts its search to the histories of files modified by the BFC.
In three of the six cases identified by \tool{} but not Simple-SZZ-Agent, the oracle BIC changed other files and was absent from the candidates.
In these cases, the methods missed the oracle BIC when generating candidates, rather than when selecting a BIC from them.

\textbf{Searches that started with code tracing sometimes stopped before investigating the actual BIC.}
A change to related code can make existing code behave incorrectly.
The commit that introduced the code changed by the fix is therefore not always the BIC.
In three cases identified by \tool{} but not AgentSZZ, AgentSZZ selected the commit that introduced the code without investigating the related changes that caused the bug.
SZZ-Agent showed a similar pattern in three cases.
It selected an initial \texttt{git blame} candidate and stopped, even though the oracle BIC was outside the candidate set.
In contrast, \tool{} investigated the related changes and identified the oracle BIC.
Even when a method can search beyond the initial candidates, it can miss the actual BIC if it stops after finding a plausible candidate through code tracing.

\summarybox{Implications}{%
	When designing SZZ methods, researchers should be aware that narrowing the search space can exclude the actual BIC.
	Searches that start with code tracing can stop before investigating the actual BIC.
	This finding supports \tool{}'s approach of reviewing a broader range of commits before examining individual candidates in detail.
}

\raggedbottom

\section{Threats to Validity}
\label{sec:threats}

\subsection{Internal Validity}

\noindent
\textbf{LLM nondeterminism.}
\tool{} and other agents interact with an LLM multiple times during an investigation.
Differences in these responses can lead to different search paths and BIC selections across runs of the same case.
We therefore ran the LLM- and agent-based SZZ methods, including \tool{}, three times per case and averaged the evaluation metrics across the three repetitions.
We compared case-level mean F1 scores using two-sided Wilcoxon signed-rank tests and applied Holm's correction for multiple comparisons.

\noindent
\textbf{Answer leakage.}
\tool{} may find information that reveals the oracle BIC in the repository and Git history.
We removed BIC hashes and other BIC-identifying information from BFC messages.
In Section~\ref{sec:commit-messages}, we also evaluated \tool{} with all commit messages removed.
We checked the logs of the main experiment and confirmed that every returned BIC hash appeared in a tool output before the final answer.
These hashes were therefore available to \tool{} during the investigation.
However, we cannot rule out the use of bug information learned during model training.

\noindent
\textbf{Dataset reconstruction.}
When \tool{}'s answers did not match the original oracle labels, we examined other relevant commits and corrected the label errors we found.
When the answers matched, we verified the labels by checking the changes made by the BFC and the oracle BIC.
However, we did not examine other commits in these matching cases.
We may therefore have missed additional BICs or ambiguity about which change introduced the bug.

\subsection{External Validity}

\noindent
\textbf{Language and project scope.}
We evaluated 79 cases from 70 Java open-source repositories.
The cases cover many projects but only one programming language.
LLMs may perform differently across languages, and our results may not generalize to other languages or proprietary projects.

\noindent
\textbf{Model and version dependence.}
We evaluated \tool{} and the LLM- and agent-based SZZ baselines using the same model, Sonnet 5.
Under these conditions, \tool{} achieved the highest BIC identification accuracy.
Different models or agent versions may change the search processes and relative accuracy of these methods.

\section{Conclusion and Lessons Learned}

To improve BIC identification accuracy, we explored search processes that do not start with code tracing.
We evaluated \tool{} without SZZ-specific tools, investigation instructions, or prepared candidate sets.
On 79 Java cases from 70 repositories, \tool{} achieved an F1 score of 0.771, significantly outperforming all evaluated baselines.
Its searches often began by reviewing commit histories and messages, followed by a closer inspection of individual diffs.

Without commit messages, \tool{} made more Bash calls, took longer, and cost more.
These increases were statistically significant, whereas the difference in F1 was not.
This suggests that commit messages help \tool{} focus its investigation on relevant changes.
Our failure analysis also found incorrect labels and cases with ambiguous BIC definitions in the Developer-Informed Oracle Dataset.
We corrected labels and excluded ambiguous or incomplete cases to create the Developer-Informed Oracle Verified dataset.
On this dataset, \tool{} achieved an F1 score of 0.938.

We summarize the lessons from Claude Code for developers and researchers.

\noindent{}
\textbf{For developers, \tool{} is a practical option for identifying BICs.}
Identifying BICs helps developers understand the cause of a bug and determine which versions need a fix.
For example, Linux kernel developers use \texttt{Fixes:} tags to record BICs, help review fixes, and determine which stable versions should receive them~\cite{linuxSubmittingPatches}.
Our study showed that \tool{} achieved high BIC identification accuracy without SZZ-specific tools or investigation instructions.
When fixing a bug, developers can use \tool{} to investigate which commit introduced it.

\smallskip{}
\noindent{}
\textbf{For researchers, SZZ methods should use semantic information about changes to find candidate commits, rather than start with code tracing.}
Our analysis showed that restricted search spaces sometimes excluded the BIC and that searches starting with code tracing sometimes stopped after finding a plausible candidate.
These findings support the search process observed in \tool{}: reviewing a broad range of commits, using commit messages to find relevant changes, and examining their diffs in detail.
However, \tool{} is a general-purpose coding agent and is not optimized specifically for BIC identification.
Future research should build on this search process to develop models or agents specialized for BIC identification, aiming to maintain high accuracy while reducing execution time and cost.

\bibliographystyle{IEEEtran}
\bibliography{bib/references}

@IEEEtranBSTCTL{IEEEBSTcontrol,
  CTLdash_repeated_names = "no"
}

@misc{anthropicClaudeCode,
  author       = {{Anthropic}},
  title        = {{Claude Code}: Overview},
  howpublished = {Claude Code Documentation},
  url          = {https://code.claude.com/docs/en/overview},
  note         = {Accessed: September 7, 2026}
}

@misc{anthropic2026sonnet5,
  author       = {{Anthropic}},
  title        = {Introducing {Claude Sonnet 5}},
  month        = jun,
  year         = {2026},
  url          = {https://www.anthropic.com/news/claude-sonnet-5},
  note         = {Accessed: October 9, 2026}
}

@article{rezk_ghost_2022,
  author  = {Christophe Rezk and Yasutaka Kamei and Shane McIntosh},
  title   = {The Ghost Commit Problem When Identifying Fix-Inducing Changes: An Empirical Study of {Apache} Projects},
  journal = {IEEE Transactions on Software Engineering},
  volume  = {48},
  number  = {9},
  pages   = {3297--3309},
  year    = {2022},
  doi     = {10.1109/TSE.2021.3087419}
}

@article{rodriguezperez2020bugs,
  author  = {Gema Rodr{\'i}guez-P{\'e}rez and Gregorio Robles and Alexander Serebrenik and Andy Zaidman and Daniel M. Germ{\'a}n and Jesus M. Gonzalez-Barahona},
  title   = {How Bugs Are Born: A Model to Identify How Bugs Are Introduced in Software Components},
  journal = {Empirical Software Engineering},
  volume  = {25},
  number  = {2},
  pages   = {1294--1340},
  year    = {2020},
  doi     = {10.1007/s10664-019-09781-y}
}

@inproceedings{watanabe2024token,
  author    = {Hiroya Watanabe and Masanari Kondo and Eunjong Choi and Osamu Mizuno},
  title     = {Benefits and Pitfalls of Token-Level {SZZ}: An Empirical Study on {OSS} Projects},
  booktitle = {Proceedings of the 31st IEEE International Conference on Software Analysis, Evolution and Reengineering},
  pages     = {776--786},
  year      = {2024},
  publisher = {IEEE},
  doi       = {10.1109/SANER60148.2024.00084}
}

@article{rani2024discussion,
  author  = {Pooja Rani and Fernando Petrulio and Alberto Bacchelli},
  title   = {On Refining the {SZZ} Algorithm with Bug Discussion Data},
  journal = {Empirical Software Engineering},
  volume  = {29},
  number  = {5},
  year    = {2024},
  note    = {Art. no. 115},
  doi     = {10.1007/s10664-024-10511-2}
}

@inproceedings{herzig2013tangled,
  author    = {Kim Herzig and Andreas Zeller},
  title     = {The Impact of Tangled Code Changes},
  booktitle = {Proceedings of the 10th International Workshop on Mining Software Repositories},
  pages     = {121--130},
  year      = {2013}
}

@article{lyu2024evaluating,
  author  = {Yunbo Lyu and Hong Jin Kang and Ratnadira Widyasari and Julia Lawall and David Lo},
  title   = {Evaluating {SZZ} Implementations: An Empirical Study on the {Linux} Kernel},
  journal = {IEEE Transactions on Software Engineering},
  volume  = {50},
  number  = {9},
  pages   = {2219--2239},
  year    = {2024},
  doi     = {10.1109/TSE.2024.3406718}
}

@inproceedings{sahal2018additions,
  author    = {Emre Sahal and Ayse Tosun},
  title     = {Identifying Bug-Inducing Changes for Code Additions},
  booktitle = {Proceedings of the 12th ACM/IEEE International Symposium on Empirical Software Engineering and Measurement},
  pages     = {1--2},
  year      = {2018},
  doi       = {10.1145/3239235.3267440}
}

@article{perezrosero2024wia,
  author        = {Salom{\'e} Perez-Rosero and Robert Dyer and Samuel W. Flint and Shane McIntosh and Witawas Srisa-an},
  title         = {{WIA-SZZ}: Work Item Aware {SZZ}},
  journal       = {Empirical Software Engineering},
  volume        = {30},
  number        = {3},
  year          = {2025},
  note          = {Art. no. 75},
  doi           = {10.1007/s10664-025-10616-2}
}

@article{kondo2026mvszz,
  author  = {Inase Kondo and Masanari Kondo and Daniel M. German and Yasutaka Kamei and Yoshiki Higo},
  title   = {{MV-SZZ}: An Empirical Study of a Majority Voting-Based {SZZ} Method},
  journal = {IEEE Transactions on Software Engineering},
  volume  = {52},
  number  = {6},
  pages   = {1911--1927},
  year    = {2026},
  doi     = {10.1109/TSE.2026.3688089}
}

@misc{linuxSubmittingPatches,
  author       = {{The Linux Kernel Project}},
  title        = {Submitting Patches: The Essential Guide to Getting Your Code into the Kernel},
  howpublished = {The Linux Kernel Documentation},
  url          = {https://cdn.kernel.org/doc/html/latest/process/submitting-patches.html},
  note         = {Accessed: September 7, 2026}
}

@inproceedings{li2024patchporting,
  author        = {Xingyu Li and Zheng Zhang and Zhiyun Qian and Trent Jaeger and Chengyu Song},
  title         = {An Investigation of Patch Porting Practices of the {Linux} Kernel Ecosystem},
  booktitle     = {Proceedings of the 21st International Conference on Mining Software Repositories},
  pages         = {63--74},
  year          = {2024},
  publisher     = {ACM},
  doi           = {10.1145/3643991.3644902}
}

@inproceedings{sliwerski2005changes,
  author    = {Jacek {\'S}liwerski and Thomas Zimmermann and Andreas Zeller},
  title     = {When Do Changes Induce Fixes?},
  booktitle = {Proceedings of the Second International Workshop on Mining Software Repositories},
  pages     = {24--28},
  year      = {2005},
  doi       = {10.1145/1083142.1083147}
}

@article{rosa2023comprehensive,
  author  = {Giovanni Rosa and Luca Pascarella and Simone Scalabrino and Rosalia Tufano and Gabriele Bavota and Michele Lanza and Rocco Oliveto},
  title   = {A Comprehensive Evaluation of {SZZ} Variants Through a Developer-Informed Oracle},
  journal = {Journal of Systems and Software},
  volume  = {202},
  year    = {2023},
  note    = {Art. no. 111729},
  doi     = {10.1016/j.jss.2023.111729}
}

@article{tang2025llm4szz,
  author  = {Lingxiao Tang and Jiakun Liu and Zhongxin Liu and Xiaohu Yang and Lingfeng Bao},
  title   = {{LLM4SZZ}: Enhancing {SZZ} Algorithm with Context-Enhanced Assessment on Large Language Models},
  journal = {Proceedings of the ACM on Software Engineering},
  volume  = {2},
  number  = {ISSTA},
  pages   = {343--365},
  year    = {2025},
  doi     = {10.1145/3728885}
}

@misc{lyu2026agentszz,
  author        = {Yunbo Lyu and Jieke Shi and Hong Jin Kang and Ratnadira Widyasari and Junda He and Yuqing Niu and Chengran Yang and Junkai Chen and Zhou Yang and Julia Lawall and David Lo},
  title         = {{AgentSZZ}: Teaching the {LLM} Agent to Play Detective with Bug-Inducing Commits},
  year          = {2026},
  eprint        = {2604.02665},
  note          = {arXiv:2604.02665v1},
  archiveprefix = {arXiv},
  primaryclass  = {cs.SE},
  doi           = {10.48550/ARXIV.2604.02665},
  url           = {https://arxiv.org/abs/2604.02665v1}
}

@inproceedings{risse2026agents,
  author        = {Niklas Risse and Marcel B{\"o}hme},
  title         = {{LLM}-Based Agents for Identifying Bug-Introducing Commits},
  booktitle     = {Proceedings of the 41st IEEE/ACM International Conference on Automated Software Engineering},
  pages         = {81--92},
  year          = {2026},
  publisher     = {ACM},
  doi           = {10.1145/3832783.3834337}
}

@misc{shi2026agenticszz,
  author        = {Yu Shi and Hao Li and Bram Adams and Ahmed E. Hassan},
  title         = {{AgenticSZZ}: Temporal Knowledge Graph-Guided Agentic Bug-Inducing Commit Identification},
  year          = {2026},
  eprint        = {2602.02934},
  note          = {arXiv:2602.02934v2},
  archiveprefix = {arXiv},
  primaryclass  = {cs.SE},
  doi           = {10.48550/ARXIV.2602.02934},
  url           = {https://arxiv.org/abs/2602.02934v2}
}

@inproceedings{xu2025swecompass,
  author  = {Jingxuan Xu and Ken Deng and Weihao Li and others},
  title   = {{SWE-Compass}: Towards Unified Evaluation of Agentic Coding Abilities for Large Language Models},
  booktitle = {Proceedings of the 43rd International Conference on Machine Learning},
  series  = {Proceedings of Machine Learning Research},
  volume  = {306},
  pages   = {140878--140902},
  year    = {2026},
  publisher = {PMLR}
}

@inproceedings{zhang2025bountybench,
  author    = {Andy Zhang and Joey Ji and Celeste Menders and others},
  title     = {{BountyBench}: Dollar Impact of {AI} Agent Attackers and Defenders on Real-World Cybersecurity Systems},
  booktitle = {Advances in Neural Information Processing Systems},
  volume    = {38},
  pages     = {190575--190694},
  year      = {2025},
  publisher = {Curran Associates, Inc.},
  doi       = {10.52202/085713-5725}
}

@article{watanabe2026agentic,
  author  = {Miku Watanabe and Hao Li and Yutaro Kashiwa and Brittany Reid and Hajimu Iida and Ahmed E. Hassan},
  title   = {On the Use of Agentic Coding: An Empirical Study of Pull Requests on {GitHub}},
  journal = {ACM Transactions on Software Engineering and Methodology},
  volume  = {35},
  number  = {10},
  pages   = {1--24},
  year    = {2026},
  doi     = {10.1145/3798166}
}

@inproceedings{merrill2026terminalbench,
  author        = {Mike A. Merrill and Alexander G. Shaw and Nicholas Carlini and others},
  title         = {{Terminal-Bench}: Benchmarking Agents on Hard, Realistic Tasks in Command Line Interfaces},
  booktitle     = {The Fourteenth International Conference on Learning Representations},
  pages         = {40903--40986},
  year          = {2026}
}

@article{kamei2013jit,
  author  = {Yasutaka Kamei and Emad Shihab and Bram Adams and Ahmed E. Hassan and Audris Mockus and Anand Sinha and Naoyasu Ubayashi},
  title   = {A Large-Scale Empirical Study of Just-in-Time Quality Assurance},
  journal = {IEEE Transactions on Software Engineering},
  volume  = {39},
  number  = {6},
  pages   = {757--773},
  year    = {2013},
  doi     = {10.1109/TSE.2012.70}
}

@inproceedings{bao2022vszz,
  author    = {Lingfeng Bao and Xin Xia and Ahmed E. Hassan and Xiaohu Yang},
  title     = {{V-SZZ}: Automatic Identification of Version Ranges Affected by {CVE} Vulnerabilities},
  booktitle = {Proceedings of the 44th International Conference on Software Engineering},
  pages     = {2352--2364},
  year      = {2022},
  publisher = {ACM},
  doi       = {10.1145/3510003.3510113}
}

@inproceedings{kim2006automatic,
  author    = {Sunghun Kim and Thomas Zimmermann and Kai Pan and Whitehead, Jr., E. James},
  title     = {Automatic Identification of Bug-Introducing Changes},
  booktitle = {Proceedings of the 21st IEEE/ACM International Conference on Automated Software Engineering},
  pages     = {81--90},
  year      = {2006},
  publisher = {IEEE},
  doi       = {10.1109/ASE.2006.23}
}

@article{davies2014comparing,
  author  = {Steven Davies and Marc Roper and Murray Wood},
  title   = {Comparing Text-Based and Dependence-Based Approaches for Determining the Origins of Bugs},
  journal = {Journal of Software: Evolution and Process},
  volume  = {26},
  number  = {1},
  pages   = {107--139},
  year    = {2014},
  doi     = {10.1002/smr.1619}
}

@article{dacosta2017framework,
  author  = {Daniel Alencar da Costa and Shane McIntosh and Weiyi Shang and Uira Kulesza and Roberta Coelho and Ahmed E. Hassan},
  title   = {A Framework for Evaluating the Results of the {SZZ} Approach for Identifying Bug-Introducing Changes},
  journal = {IEEE Transactions on Software Engineering},
  volume  = {43},
  number  = {7},
  pages   = {641--657},
  year    = {2017},
  doi     = {10.1109/TSE.2016.2616306}
}

@inproceedings{neto2018impact,
  author    = {Edmilson Campos Neto and Daniel Alencar da Costa and Uira Kulesza},
  title     = {The Impact of Refactoring Changes on the {SZZ} Algorithm: An Empirical Study},
  booktitle = {Proceedings of the 25th IEEE International Conference on Software Analysis, Evolution and Reengineering},
  pages     = {380--390},
  year      = {2018},
  publisher = {IEEE},
  doi       = {10.1109/SANER.2018.8330225}
}

@inproceedings{tang2023neural,
  author    = {Lingxiao Tang and Lingfeng Bao and Xin Xia and Zhongdong Huang},
  title     = {Neural {SZZ} Algorithm},
  booktitle = {Proceedings of the 38th IEEE/ACM International Conference on Automated Software Engineering},
  pages     = {1024--1035},
  year      = {2023},
  publisher = {IEEE},
  doi       = {10.1109/ASE56229.2023.00037}
}

@article{kim2008classifying,
  author  = {Sunghun Kim and Whitehead, Jr., E. James and Yi Zhang},
  title   = {Classifying Software Changes: Clean or Buggy?},
  journal = {IEEE Transactions on Software Engineering},
  volume  = {34},
  number  = {2},
  pages   = {181--196},
  year    = {2008},
  doi     = {10.1109/TSE.2007.70773}
}

@article{bernardi2018communication,
  author  = {Mario Luca Bernardi and Gerardo Canfora and Di Lucca, Giuseppe A. and Di Penta, Massimiliano and Damiano Distante},
  title   = {The Relation Between Developers' Communication and Fix-Inducing Changes: An Empirical Study},
  journal = {Journal of Systems and Software},
  volume  = {140},
  pages   = {111--125},
  year    = {2018},
  doi     = {10.1016/j.jss.2018.02.065}
}

@article{mcintosh2018moving,
  author  = {Shane McIntosh and Yasutaka Kamei},
  title   = {Are Fix-Inducing Changes a Moving Target? A Longitudinal Case Study of Just-in-Time Defect Prediction},
  journal = {IEEE Transactions on Software Engineering},
  volume  = {44},
  number  = {5},
  pages   = {412--428},
  year    = {2018},
  doi     = {10.1109/TSE.2017.2693980}
}

@inproceedings{pornprasit2021jitline,
  author    = {Chanathip Pornprasit and Chakkrit Tantithamthavorn},
  title     = {{JITLine}: A Simpler, Better, Faster, Finer-grained Just-in-Time Defect Prediction},
  booktitle = {Proceedings of the 18th International Conference on Mining Software Repositories},
  pages     = {369--379},
  year      = {2021},
  publisher = {IEEE},
  doi       = {10.1109/MSR52588.2021.00049}
}

@article{vandehei2021leveraging,
  author  = {Bailey Vandehei and Daniel Alencar da Costa and Davide Falessi},
  title   = {Leveraging the Defects Life Cycle to Label Affected Versions and Defective Classes},
  journal = {ACM Transactions on Software Engineering and Methodology},
  volume  = {30},
  number  = {2},
  pages   = {24:1--24:35},
  year    = {2021},
  doi     = {10.1145/3433928}
}

@book{manning2008introduction,
  author    = {Christopher D. Manning and Prabhakar Raghavan and Hinrich Sch{\"u}tze},
  title     = {Introduction to Information Retrieval},
  publisher = {Cambridge University Press},
  year      = {2008},
  isbn      = {0521865719}
}

@article{Wilcoxon1945Biometrics,
  author  = {Frank Wilcoxon},
  title   = {Individual Comparisons by Ranking Methods},
  journal = {Biometrics Bulletin},
  volume  = {1},
  number  = {6},
  pages   = {80--83},
  year    = {1945}
}

@article{wen2022quick,
  author  = {Fengcai Wen and Csaba Nagy and Michele Lanza and Gabriele Bavota},
  title   = {Quick remedy commits and their impact on mining software repositories},
  journal = {Empirical Software Engineering},
  volume  = {27},
  number  = {1},
  year    = {2022},
  note    = {Art. no. 14},
  doi     = {10.1007/s10664-021-10051-z}
}

@article{holm1979simple,
  author  = {Sture Holm},
  title   = {A Simple Sequentially Rejective Multiple Test Procedure},
  journal = {Scandinavian Journal of Statistics},
  volume  = {6},
  number  = {2},
  pages   = {65--70},
  year    = {1979}
}

@inproceedings{wen2019exploring,
  author    = {Ming Wen and Rongxin Wu and Yepang Liu and Yongqiang Tian and Xuan Xie and Shing-Chi Cheung and Zhendong Su},
  title     = {Exploring and Exploiting the Correlations between Bug-Inducing and Bug-Fixing Commits},
  booktitle = {Proceedings of the 27th ACM Joint Meeting on European Software Engineering Conference and Symposium on the Foundations of Software Engineering},
  pages     = {326--337},
  year      = {2019},
  publisher = {ACM},
  doi       = {10.1145/3338906.3338962}
}

@inproceedings{tian2022good,
  author    = {Yingchen Tian and Yuxia Zhang and Klaas-Jan Stol and Lin Jiang and Hui Liu},
  title     = {What Makes a Good Commit Message?},
  booktitle = {Proceedings of the 44th International Conference on Software Engineering},
  pages     = {2389--2401},
  year      = {2022},
  publisher = {ACM},
  doi       = {10.1145/3510003.3510205}
}

@inproceedings{li2023commit,
  author    = {Jiawei Li and Iftekhar Ahmed},
  title     = {Commit Message Matters: Investigating Impact and Evolution of Commit Message Quality},
  booktitle = {Proceedings of the 45th IEEE/ACM International Conference on Software Engineering},
  pages     = {806--817},
  year      = {2023},
  publisher = {IEEE},
  doi       = {10.1109/ICSE48619.2023.00076}
}

@inproceedings{dhaouadi2024rationale,
  author    = {Mouna Dhaouadi and Bentley James Oakes and Michalis Famelis},
  title     = {Rationale Dataset and Analysis for the Commit Messages of the {Linux} Kernel Out-of-Memory Killer},
  booktitle = {Proceedings of the 32nd IEEE/ACM International Conference on Program Comprehension},
  pages     = {415--425},
  year      = {2024},
  publisher = {ACM},
  doi       = {10.1145/3643916.3644413}
}

\end{document}